\documentstyle[11pt,epsf]{article}
\input epsf.tex
\begin{document}
\title{Positronium Hyperfine Splitting in Non-commutative Space
at the Order $\alpha^6$}
\vspace{22pt}
\author{M. Haghighat$^{a,c}$\thanks{E-mail: mansour@cc.iut.ac.ir},\ \ S.M.
Zebarjad$^{b,c}$\thanks{E-mail: zebarjad@physics.susc.ac.ir}\ \ and\ \
F. Loran$^{a,c}$\thanks{E-mail: farhang@theory.ipm.ac.ir }\\
{\it $^a$Physics Department, Isfahan University of Technology (IUT),
Isfahan 84154, Iran}, \\
{\it $^b$Physics Department,  Shiraz University,
 Shiraz 71454,  Iran,} \\
{\it $^c$Institute for Studies in Theoretical Physics and Mathematics (IPM), Tehran
19395, Iran. }}
\date{ }
\maketitle
\begin{abstract}
We obtain positronium Hyperfine Splitting owing to the non-commutativity of space and
show that, in the leading order, it is proportional to $\theta \alpha^6$ where,
$\theta$ is the parameter of non-commutativity. It is also shown that spatial
non-commutativity splits the spacing between $n=2$ triplet excited levels
$E(2^3S_1)\rightarrow E(2^3P_2)$ which provides an experimental test on the
non-commutativity of space.
\end{abstract}
\newpage
\section{Introduction}
The question of measuring of the spatial non-commutativity
effects, in physical processes, is under intensive interest.
Non-commutative QED (NCQED) seems to be a straightforward method
to examine such effects. For this purpose, one needs a precise
experimental data such as  positronium hyperfine splitting (HFS)
among the other processes. The basic difference between NCQED and
QED is the existence of new interactions (3-photon and 4-photon
vertices) which complicate the calculations in NCQED. Although the
Feynman rules of this theory are given in \cite{a1,a2}, to
apply these rules to bound state, one needs special treatments
like Bethe-Salpeter (BS) approach \cite{a3} or non-relativistic
QED (NRQED)\cite{a4}.  In our preceding letter \cite{a41}, using BS
equation, we have shown that up to the order $\alpha^4$ no
spin-dependent correction owing to the spatial non-commutativity
appears in the positronium spectrum.  Therefore one should
calculate the higher order corrections.  In this letter we
calculate the corrections to the positronium by using NRQED
method.  In section 2, we introduce NRQED-vertices in the
NC-space. Consequently, in section 3, we use the modified NRQED to
determine  HFS at the lowest order. In this section, we show that
our calculations at the leading order lead to the corrections at
the order of $\theta \alpha^6$, where $\theta$ is the parameter of
non-commutativity. At the end, we summarize our results.

\section{NRQED in non-commutative space}
NRQED is an effective field theory which simplifies the bound state calculation. To
apply this technique in non-commutative space one should modify the NRQED vertices by
performing $\frac{p}{m_e}$ expansion on NCQED scattering amplitude. In doing so, one
obtains an effective theory of non-relativistic particles which permits the direct
application of well tested techniques based on Schr\"{o}dinger's equation.  Now,
comparing NCQED scattering amplitudes with NRQED can completely determine the
matching coefficients.  Some of the
 vertices with their appropriate matching coefficients, are shown in Fig. 1.
They contribute to the tree level matching to get the leading order bound state
energy shift.  One should note that these coefficients apart from a phase factor are
very similar to the standard NRQED \cite{a411,a412}.  This similarity is owing to the
fact that the scattering amplitude of $e^+e^-$ in NCQED is independent of the
parameter of non-commutativity of space \cite{a42}.  The other vertices which are not
shown in Fig. 1 and have not counterpart in the standard NRQED, due to the existence
of the three and four photon vertices have contributions to higher order corrections
to energy shift. Now, by using the first graph of Fig. 1 and expanding the vertices
up to order $\theta$, one can easily verify the results of refs.\cite{a41,a141} at the
order $\theta\alpha^4$ as
\begin{equation}
\Delta E=\left<\alpha\frac{\Theta.{\bf
L}}{r^3}\right>=\theta\alpha^4\frac{P_{n,l}}{l(l+\frac{1}{2})(l+1)},
\end{equation}
Such an energy shift is spin-independent and therefore has not any contribution to
HFS.
\begin{figure}
\centerline{\epsfxsize=4in\epsffile{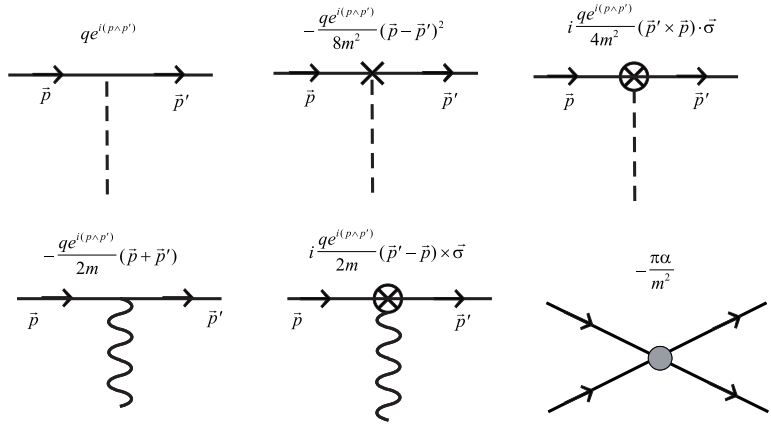}} \caption{NRQED
vertices in non-commutative space. } \label{a6a}
\end{figure}
\section{Positronium HFS at the leading order}
By using the modified NRQED we can determine the diagrams which
contribute to the lowest order of HFS (Fig. 2).
\begin{figure}
\centerline{\epsfxsize=4in\epsffile{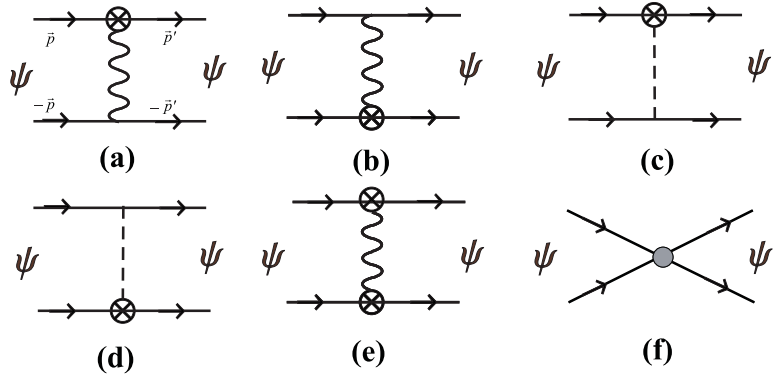}} \caption{All the
bound state diagrams at LO.  } \label{a6b}
\end{figure}
We can now calculate each diagram separately as follows:
\begin{eqnarray}\label{p1}
\triangle E_a=\int \frac{d^3p d^3p' }{(2\pi)^6}\psi^*({\bf p}') \Gamma_a({\bf p},{\bf
p}')
 \psi ({\bf p}),
\end{eqnarray}
with
\begin{eqnarray}\label{p2}
\Gamma_a({\bf p},{\bf p}')&=&\left[\frac{-ie({\bf p}'-{\bf p})
\times{{\bf{\sigma_{1}}}}}{2m_e}e^{i p\wedge p'}\right]_i
\frac{-1}{({\bf p}-{\bf p}')^2}\nonumber\\& &
\left[\delta_{ij}-\frac{({\bf p}-{\bf p}')_i({\bf p}-{\bf
p}')_j}{({\bf p}-{\bf p}')^2}\right] \left[\frac{e({\bf p}'+{\bf
p})}{2m_e}e^{i p\wedge p'}\right]_j,
\end{eqnarray}
where ${\bf p}\wedge {\bf p}'=\frac{1}{2}\theta_{\mu\nu}p_\mu p'_
\nu$ and $\theta_{\mu\nu}$, the parameter of the non-commutativity
is given as
\begin{eqnarray}\label{a3}
\theta_{\mu\nu}=i\left[x_\mu ,x_\nu\right].
\end{eqnarray}
It is shown that $\theta_{0i}\ne 0$ leads to some problems
 with the unitarity of field theory and the concept of causality \cite{a43,a44},
therefore in our calculations we consider $\theta_{0i}= 0$.\\
After some algebra Eq.(\ref{p1}) yields
\begin{eqnarray}\label{p4}
\triangle E_a&=&\frac{i e^2}{2m_e^2}\int \frac{d^3 p d^3 p'}{(2\pi)^6} \psi^*({\bf
p}')\frac{{\bf{\sigma_1}}.{\bf p}\times{\bf p}'}{({\bf p}-{\bf p}')^2}
e^{i\theta_{ij}p_ip'_j}\psi({\bf p})\nonumber\\
&=&\frac{e^2}{8\pi m_e^2}\int d^3r\left[\psi^*\left({\bf r}+i\theta.\nabla\right)
\frac{{\bf r}\times{\bf p}}{r^3}.{\bf{\sigma_1}}\right]\psi({\bf r})\nonumber\\
&=&\frac{\alpha}{2m_e^2}\left<\frac{{\bf S}_1.{\bf L}} {r^3}\right>-\frac{3\alpha}
{m_e^2}\int d^3 r (\Theta .{\bf L} \psi^*)\frac{{\bf S}_1.{\bf
L}}{r^5}{\psi}+O(\alpha^7),
\end{eqnarray}
where $(\theta.\nabla)_i=\theta_{ij}\partial_j$ and $
\Theta=(\theta_{23}, \theta_{31}, \theta_{12})$. In the third
equality we used
\begin{eqnarray}\label{p5}
\psi^*({\bf r}+i\theta .\nabla)=\psi^*({\bf r})+i(\nabla \psi^*({\bf
r}).\theta.\nabla) +O(\theta^2).
\end{eqnarray}
One should note that the first term in Eq.(\ref{p4}) is the usual term in NRQED which
is of the order $\alpha ^4$. But the second term which is appeared in Eq.(\ref{p4}),
owing to the spatial non-commutativity is of the order $\theta \alpha^6$. Nonexistence
of the terms at the order of $\alpha^4$ which carry $\theta$-dependence is a
remarkable result which happens in all diagrams of HFS. Indeed this fact is due to
appearance of $\psi^*({\bf r}+i\theta.\nabla)$ instead of $\psi^*({\bf r})$ in all
energy-correction expressions. Therefore, to obtain the energy corrections for HFS at
the order $\alpha^6$ one should once calculate the corrections up to the lowest order
of NRQED (i.e. Fig. 2). In the other words, the $\alpha ^6$-corrections in NRQED
calculations of commutative space lead to the higher order of
$\alpha$ in non-commutative space. \\
Now we work out the Figs. 2(b-f) as follows:
\begin{eqnarray}\label{p6}
\triangle E_b&=&\triangle E_a({\bf S}_1\rightarrow {\bf S}_2)\nonumber\\
\triangle E_c+\triangle E_d&=&\frac{1}{2}(\triangle E_a+\triangle
E_b)
\end{eqnarray}
\begin{eqnarray}\label{p7}
\triangle E_e=\int \frac{d^3p d^3p' }{(2\pi)^6}\psi^*({\bf p}')\Gamma_e({\bf p},{\bf
p}')
 \psi ({\bf p}),
\end{eqnarray}
with
\begin{eqnarray}\label{p8}
\Gamma_e({\bf p},{\bf p}')&=&\left[\frac{-ie({\bf p}'-{\bf
p})\times{\bf{\sigma_1}}}{2m_e}e^{i p\wedge p'}\right]_i \frac{-1}{({\bf
p}-{\bf p}')^2}\nonumber\\& & \left[\delta_{ij}-\frac{({\bf
p}-{\bf p}')_i({\bf p}-{\bf p}')_j}{({\bf p}-{\bf p}')^2}\right]
\left[\frac{-ie({\bf p}'-{\bf p})\times{\bf{\sigma_2}}}{2m_e}e^{i p\wedge
p'}\right]_j,
\end{eqnarray}
which results in
\begin{eqnarray}\label{p9}
\triangle E_e&=&\frac{e^2}{4m_e^2}\int d^3r\psi({\bf r})\psi^*({\bf r}+i\theta.\nabla)
\left[-{\bf{\sigma_1}}.{\bf{\sigma_2}}\nabla^2+({\bf{\sigma_1}}.\nabla)({\bf{\sigma_2}}.\nabla)\right]\frac{1}{4\pi
r}\nonumber\\ &=&(\ldots)+\frac{3e^2}{16\pi m_e^2}\int d^3r\psi({\bf
r})\tilde\Gamma_e\psi^*({\bf r}),
\end{eqnarray}
where $(\ldots)$ means the usual part of the energy shift and
\begin{eqnarray}\label{p10}
\tilde\Gamma_e=\left[\frac{{\bf{\sigma_1}}.{\bf{\sigma_2}}}{r^5}\Theta.{\bf
L}-\frac{{\bf{\sigma_2}}.{\bf r}}{r^5} {\bf{\sigma_1}}.(\Theta\times {\bf
\hat{p}})-\frac{{\bf{\sigma_1}}.{\bf r}}{r^5}{\bf{\sigma_2}}.(\Theta\times {\bf
\hat{p}})-\frac{5} {r^7}({\bf{\sigma_1}}.{\bf r})({\bf{\sigma_2}}.{\bf r})\Theta.{\bf
L}\right],
\end{eqnarray}
where ${\bf \hat{p}}=-i\nabla$. The final diagram (Fig. 2f) has not any contribution
at the order of our interest. For $S=1$ one can easily find
\begin{eqnarray}\label{p11}
\triangle E^{\hbox{NC}}_a+\triangle E^{\hbox{NC}}_b&=&\frac{-3e^2}{4\pi m_e^2}\int d^3
r \left[\frac{\Theta.{\bf L}}
{r^5}\psi^*({\bf r})\right]\ell\psi({\bf r})\nonumber\\
\triangle E^{\hbox{NC}}_a+\triangle E^{\hbox{NC}}_b&=&\frac{1}{2}(\triangle E^{\hbox{NC}}_c+
\triangle E^{\hbox{NC}}_d)\nonumber\\
\triangle E^{\hbox{NC}}_e&=&\frac{3e^2}{16\pi m_e^2}\int d^3 r \psi({\bf
r})\overline{\Gamma}\psi^*({\bf r})
\end{eqnarray}
where
\begin{eqnarray}\label{p12}
\overline{\Gamma}=\frac{\Theta.{\bf L}}{r^5}-\frac{2}{r^5}\left\{\begin{array}{l}
z(\Theta\times {\bf{\hat{p}}})_3\\{\bf r}.(\Theta\times{\bf{
\hat{p}}})-2z(\Theta\times{\bf{ \hat{p}}})_3\\z (\Theta\times
{\bf{\hat{p}}})_3\end{array}\right\}-\frac{5}{r^7}\left\{\begin{array}{l}
z^2\\r^2-2z^2\\z^2\end{array}\right\}\Theta.{\bf L}.
\end{eqnarray}
The superscript NC in Eq. (\ref{p11}) means the non-commutative part of the energy
shift and three lines in the Eq.(\ref{p12}) are related to $S_z=1,0,-1$,
respectively. Meanwhile for the spin zero
state ($S=0$), all contributions to the energy shift are zero.\\
The average of $\triangle E_e$ over the triplet is zero, which
means the spin-spin interaction part carries no correction in
average and therefore the hyperfine splitting due to
the non-commutativity becomes
\begin{eqnarray}\label{p13}
\delta E^{\hbox{NC}}=\frac{9e^2}{8\pi m_e^2}\int d^3r\left[\frac{\Theta.{\bf
L}}{r^5}\psi^*_{nlm}({\bf r})\right]\ell\psi_{nlm}({\bf r}),
\end{eqnarray}
where $\psi_{nlm}$ is the wave function of the positronium in the commutative space
with the Coulomb potential and we have defined $\delta E^{\hbox{NC}}=\triangle
E^{\hbox{NC}}(S=1)-\triangle E^{\hbox{NC}}(S=0)$.  If the $z$-axis is chosen parallel
to the vector $\Theta$, the above result simplifies into
\begin{eqnarray}\label{p14}
\delta E^{\hbox{NC}}=\frac{9e^2}{8\pi m_e^2}\left|\Theta\right|\ell m
\left<\frac{1}{r^5}\right> ={\frac
{\left|\Theta\right|}{{\lambda}_e^2}}\alpha^6m_e\ell m f(n,l),
\end{eqnarray}
where $\lambda_e$ is the Compton wave length of the electron and $f(n,l)$ is defined
as
\begin{equation}
f(n,l)=\frac{P^{(1)}_{n,l}}{l(l+\frac{1}{2})(l+1)(l+\frac{3}{2})(l+2)}+
\frac{P^{(2)}_{n,l}}{(l-1)(l-\frac{1}{2})l(l+\frac{1}{2})(l+1)}.
\end{equation}
One should note that the divergence of $\delta E^{\hbox{NC}}$ at $l=1$ is owing to
singularity of $\left<\frac{1}{r^5}\right>$ at $r=0$, the region where
$\theta$-expansion is not well-defined.  Actually, it is shown that $\theta$-expanded
NCQED is not renormalizable \cite{a10}. \par The $\theta$ expansion imply a cut-off
$\Lambda\sim\frac{1}{\sqrt{\left|\Theta\right|}}$ while the validity of NRQED requires
$p \le m_e=\frac{1}{\lambda_e}$. Since $\sqrt{\left|\Theta\right|}\le \lambda_e$, the
appropriate cut-off is $\Lambda=\frac{1}{\lambda_e}$. Therefore the energy shift for
$n=2$, $l=1$ can be obtained as
\begin{equation}
\delta
E^{\hbox{NC}}=\frac{3}{512}m_e\left(\frac{\left|\Theta\right|}{\lambda_e^2}\right)
\left[\ln2-\gamma-\ln\alpha\right]\alpha^6.
\end{equation}
\par
The above result should be added to the values of HFS derived in NRQED at the order
$\alpha^6$. The reported uncertainties on the experimental values of
$E(2^3S_1)\rightarrow E(2^3P_2)$ are about $0.1$ MHz \cite{a11}, that give an upper
bound ${\frac {\left|\Theta\right|}{{\lambda}_e^2}}\sim 10^{-1}$. Therefore
determining the value of $\left|\Theta\right|$ requires more accurate experiments.
\section{Summary}
Using NRQED method in the non-commutative space, we have obtained that there is not
any correction at the order $\alpha^4$ for the HFS of positronium, the order
$\alpha^4$ corrections are spin independent. The correction to the energy shift is
started at the order $\alpha^6$, Eqs. (\ref{p13}-\ref{p14}), and it depends on $\ell$
and $m$ quantum numbers. Therefore it dosen't have any contribution to the
$E(1^3S_1)\rightarrow E(1^1S_0)$ (in the spectroscopic notation $n^{2S+1}L_j$), while
for $\ell\neq 0$ there is $2\ell +1$ different shifts. Consequently, a closer look at
the spacing between $n=2$ triplet excited levels ($E(2^3S_1)\rightarrow E(2^3P_2)$),
which has already been measured [14-17] can provide an experimental test on
the non-commutativity of space.\\
{\bf Acknowledgement}\\
 S.M.Z.  gratefully acknowledges research funding from Shiraz university and IPM.
 The research of
 M.H. and F.L. was supported by Isfahan University of Technology (IUT) and IPM.

\end{document}